# PIEEG: Turn a Raspberry Pi into a Brain-Computer-Interface to measure biosignals

Ildar Rakhmatulin* – PhD electronic researcher
Sebastian Völkl - Brain-Computer-Interface developer

**Abstract**
This paper presents an inexpensive, high-precision, but at the same time, easy-to-maintain PIEEG board to convert a RaspberryPI to a Brain-computer interface. This shield allows measuring and processing eight real-time EEG(Electroencephalography) signals. We used the most popular programming languages - C, C++ and Python to read the signals, recorded by the device . The process of reading EEG signals was demonstrated as completely and clearly as possible. This device can be easily used for machine learning enthusiasts to create projects for controlling robots and mechanical limbs using the power of thought. We will post use cases on GitHub (https://github.com/Ildaron/EEGwithRaspberryPI) for controlling a robotic machine, unmanned aerial vehicle, and more just using the power of thought.

*email: ildarr2016@gmail.com

**Abbreviation**

| | |
|---|---|
| BCI | Brain-computer interface |
| EEG | Electroencephalogram |
| SBC | Single-board computer |
| IC | Integrated circuit |
| ADC | Analog digital converter |

## 1. Introduction

The popularity of BCI's has increased significantly in recent years, making neuroscience more accessible to scientists and engineers from various fields of activity.Neuroscience in particular has not stood aside, and today machine learning is widely used to determine the relationship between thoughts and EEG. We believe that the relationship between machine learning and neuroscience will only grow as the scope of BCI expands significantly. Today BCI's are actively used not only for medical purposes [1, 2020] but also in many other industries, such as videogames [2, 2020], and this direction is auspicious because it's one of the fastest-growing [3, 2016]. As for other fields, Vo et al [4, 2019] and Rakhmatulin [5, 2020] have providedan overview of the possibilities of machine learning and EEG and described the prospects in this area quite favorably.

In recent years, many devices were introduced to the market as low-cost devices, but their prices also usually started from a thousand dollars.At the same time, there is a problem in evaluating the quality of the proposed devices on the market. In the review paper, Rashid et al. [6, 2020] showed that BCI developers use different evaluation metrics to confirm the effectiveness of the developed devices, which complicates the analysis of these BCIs. To avoid this problem, in addition to the technical features, we presented the results of the experiments as completely as possible on the GitHub page and uploaded the datasets for different cases.

In recent years, the activity on the BCI market has inspired optimism and gave reason to hope that BCI may soon can become an everyday means of human-computer interaction. However the positive start in the field of BCI hardware was interrupted by chip shortage [7, 2021], which complicated the production process, drove up the cost of ICs and made them scarce. Therefore, it has become expensive to develop full-fledged devices [8, 2021]. For this reason, we have developed an Open-Source board that is designed to convert a RaspberryPI to a Brain-computer interface. We made a shield because it reduced the number of ICs on the board and increased reliability. We chose the RaspberryPi, mainly because it is the most popular SBC on the market [9, 2017]. It should also be noted that one of the limiting factors for the widespread use of BCI is the difficulty in reading EEG signals, as they are susceptible to electromagnetic, motion, and other types of noise. [10, 2021]. Therefore, signal processing by software is just as important as the hardware component of the BCI. We have developed an open-source software where data collection takes place in C, which is necessary for high speed, and signal processing in real-time by Python. We chose Python because it is one of the most popular languages in data science [11, 2015] with many forums and documentation.

**1.1 Review the paper**
The following papers presented devices which in their characteristics are close to our device. Tyler et al. [12, 2015] presented a board for measuring biosignals. The authors didn't provide information about the component manufacturer, PCB layout, or PCB specifications. For these reasons, it is very difficult to judge the potential of the device. Uktveris et al [13, 2018] developed a board for reading biosignals with ADC ADS1299. The board presented in their work is aimed at the Arduino market as a shield for the MEGA2560 models. Therefore, this board is only a transmitter, and signal processing must be done on a computer. As for the combination of single-board computers and an ADC for reading bio signals, we did not find similar devices among the published works. We found the most similar project in GitHub (https://github.com/wjcroft/RaspberryPiADS1299) - but the publication on this work is not presented, so it is very difficult to understand the quality of the received EEG signal. For example, it is not clear whether voltage regulators were used, without which the ADS 1299 would not be able to provide a clean EEG signal. We also did not find software for signal processing. The development of the hardware part is an applied science and from the engineering point of view, Gerber files, BOM files, and electrical circuits are needed to confirm the experiments.. Many works are stated as open sources, but the authors did not indicate the source, so their statements about the openness of the sources can be considered advertising. As for the SBCs for reading signals, we found papers where SBCs were at best used as transmitters for sending data to a desktop computer [14, 2016], [15, 2021], but the capabilities of the SBC are greater than just being a transmitter. For these reasons, we developed a shield that receives data and transmits it to the RaspberryPi, and we have written software that processes data in real-time on RaspberryPi. Thus, we can say that the RaspberryPi has become a BCI. A general view of the device showed in fig. 1.

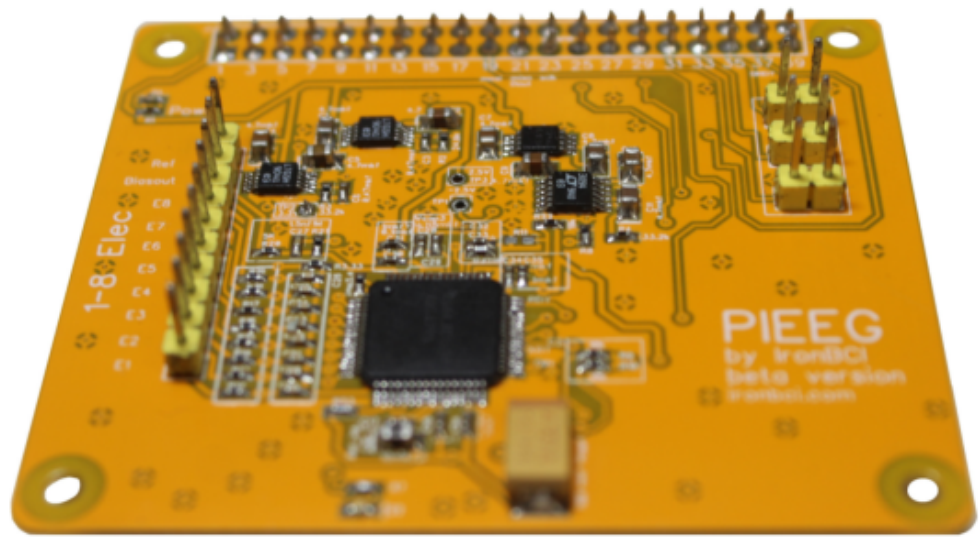

Fig.1. General view of the device

The possibilities of the device are very extensive and largely depend on the user how he wants to use this device. To start the measurement process, electrodes must be connected to a device that reads data, processes received signals and controls external devices, fig.2.

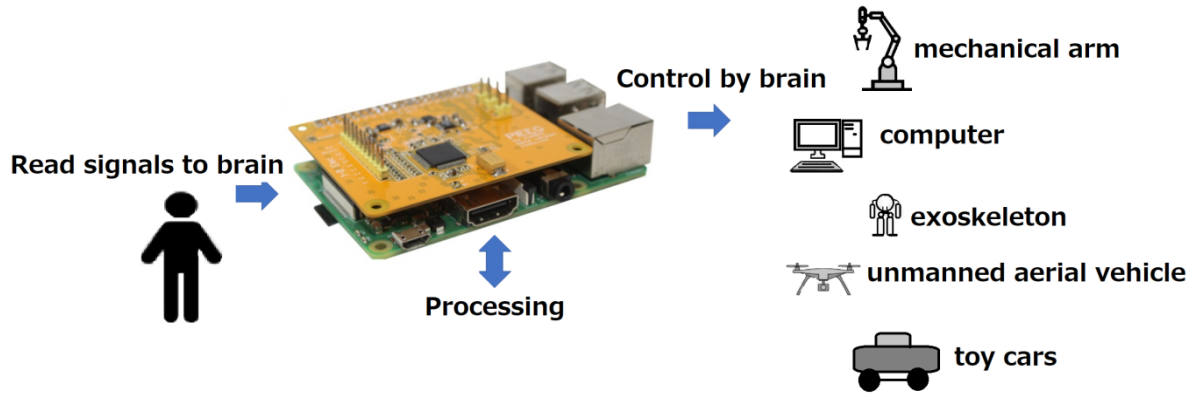

Fig. 2. Schematic representation of the use of the device

This device can be used as a shield for OrangePI (http://www.orangepi.org/), BananaPI (https://www.banana-pi.org/), and some other single-boards computers, for this just need to make a match on GPIO40.

## 2. Technical details
Table 1 lists the main technical data of the device

Table 1. PIEEG characteristic

|    | Description | Characteristics |
|----|-------------|-----------------|
| 1  | Channels    | 8 channels for connecting electrodes, 1 channel for connecting a reference, and 1 channel for connecting a bias signal, with common-mode noise suppression |
| 2  | Protocol    | Data transfer via SPI with a frequency from 250 SPS to 16 kSPS |
| 3  | Resolution  | 24 bits |
| 4  | Gain signal | 1, 2, 4, 6, 8, 12, 24 |
| 5  | Impedance   | Control connections |
| 6  | Common-Mode Rejection Ratio CMRR | 120 |
| 7  | Internal noise | 0.4 µV |
| 8  | External noise | 0.8 µV |
| 9  | Noise/Ration SNR | 130 dB |
| 10 | Indication status | Of power and live bits for ADS1299 |
| 11 | User switch | 3 |

The free 33 PIN from RaspberryPI GPIO40 can be used for user tasks, for example - connecting any external devices.
For the measurement process, the PIEEG should be connected to the Raspberry PI3 or RaspberryPI4. Then the device must be connected to a battery (power supply) and linked to the electrodes. **Full**

**galvanic isolation from the mains required.** This also applies to the monitor. Only use a monitor that is powered by the same battery as and for the RaspberryPI, as in the fig.3-a. Electrodes should be positioned according to International 10-20 system, fig.3-b. As electrodes we used dry electrodes, the advantages of which are described in this article [16, 2021].

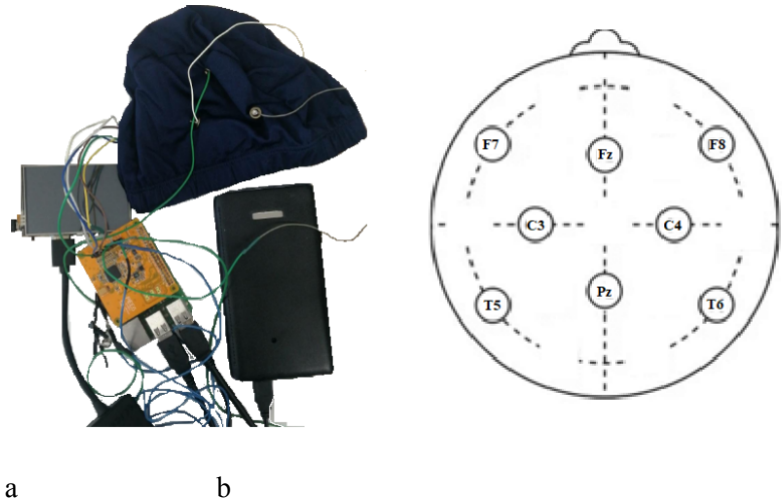

a               b

Fig. 3. Assembled device. Screen, battery, cap and 8 electrolytes connected by according to International 10-20 system -a, location of electrodes - b

## 3. Software

Today many signals processing software, both commercial and non-commercial can be used for signal processing tasks. The following open source programs are on the market: Brainstorm (https://neuroimage.usc.edu/brainstorm/Introduction), TAPEEG (https://sites.google.com/site/tapeeg/), sLORETA (https://neurofeedbackalliance.org/sloreta/), Brainflow (https://brainflow.org/), Timeflux (https://timeflux.io/). But it is worth noting, that the software review is just as complicated by the lack of any comparison criteria as for the hardware, as we wrote earlier. At the same time, we do not share the policy in which the authors of the hardware part need to add their devices to the software. Since this policy increases the size of the software and the compilation time,t is much easier to make and maintain a simple and understandable protocol for data transfer. For example, in our case, for the data to the software is transmitted by an array for 8 channels, fig.4.

```
RESTART: /home/pi/Desktop/real_time.py
library ok
[[1665846   328   2118 1407067 2256662 2162659 1274185  751494]
 [1615339   532     92 1357237 2205648 2111584 1224448  702049]
 [1670022   344   2165 1411160 2260884 2166896 1278270  755612]
 [1671682   310   2571 1412657 2262494 2168555 1279801  757219]
 [1666279   328   2117 1407489 2257074 2163077 1274607  751929]
 [1615494   532     96 1357382 2205782 2111721 1224592  702203]
 [1629223   504    356 1371033 2219650 2125581 1238215  715754]
 [1669696   346   2109 1410839 2260548 2166560 1277941  755285]
 [1671703   312   2546 1412668 2262501 2168567 1279814  757248]
 [1668234   312   2224 1409360 2259030 2165049 1276476  753829]
 [1622883   508    325 1364689 2213227 2119170 1231888  709460]
 [1629738   511    325 1371548 2220130 2126064 1238734  716294]
 [1669564   349   2104 1410705 2260393 2166403 1277807  755159]
     1      2      3       4       5       6       7       8
```

Fig. 4. Example of data transfer from device to software

The software should be as clear and simple as possible. When changing the code and using external libraries, it is important to have as little compilation effort as possible. We did not find any suitable

software and developed our own software. Software details can be found on the GitHub page
https://github.com/Ildaron/EEGwithRaspberryPI/tree/master/GUI

**4. Technical tests**

In addition to the technical details presented in Table 1, we made a typical check of the device for artifacts. The chewing artifact is one of the most pronounced artifacts, Wei, et al. [17, 2016] described this artifact in detail in the EEG signals. Unlike chewing, blinking artifact [18, 2009] is quite often used to control external objects, for example, to control electric wheelchairs [19, 2010]. For PIEEG chewing and blinking artifacts detection in real-time presented in fig.5.

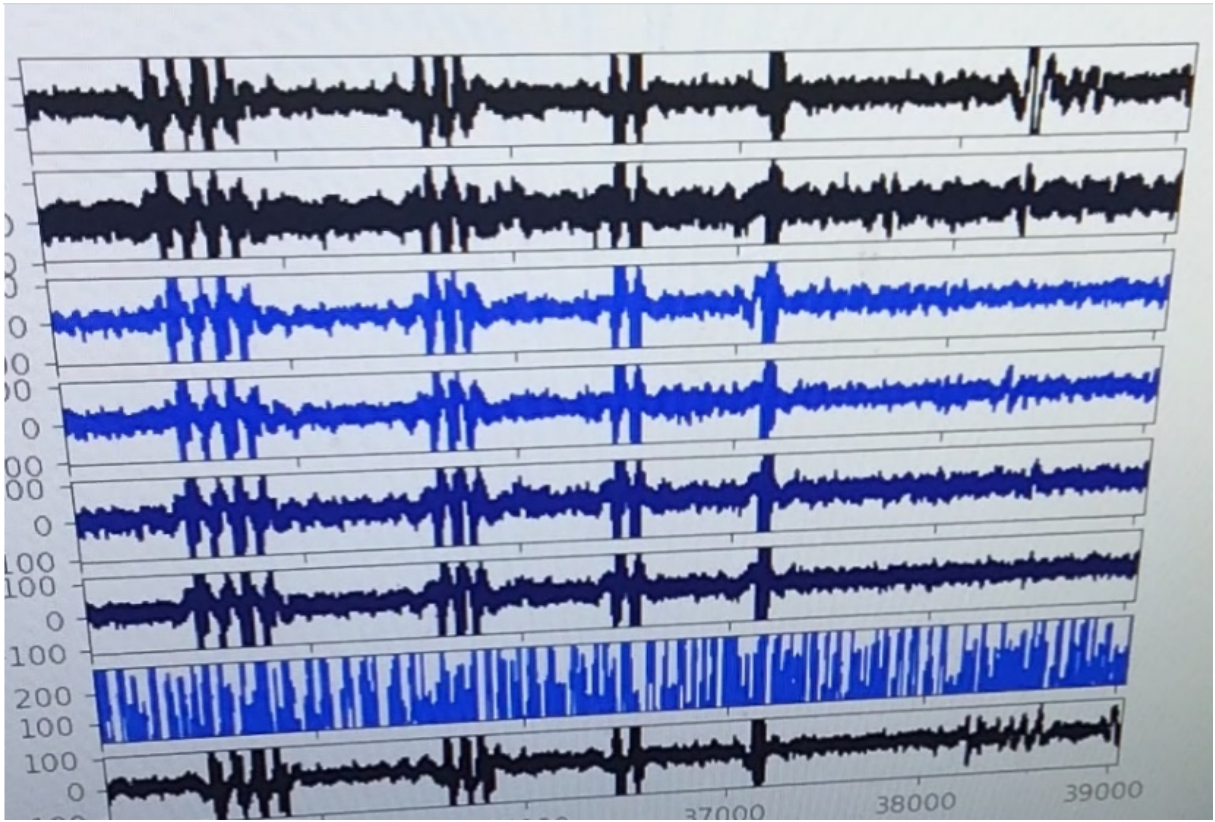

Fig.5. Chewing and blinking with EEG signal on PIEEG **(real-time)**

Band–pass filter with 1-30 Hz for the same data (**not real-time**) in fig. 8.

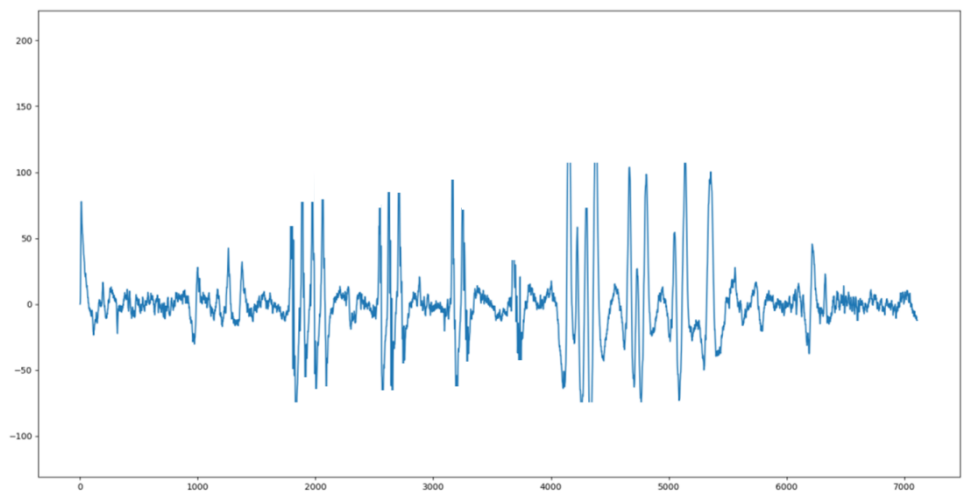

Fig. 6. Pass filter with 1-30 Hz for artifact and blinking

Alpha rhythm of alpha waves is the rhythmic activity of the EEG, recorded in the primary with open or closed eyes at rest in 85-95%. This phenomenon is well studied and described in many works [20, 2019]. The alpha wave detection process shown in fig. 9.

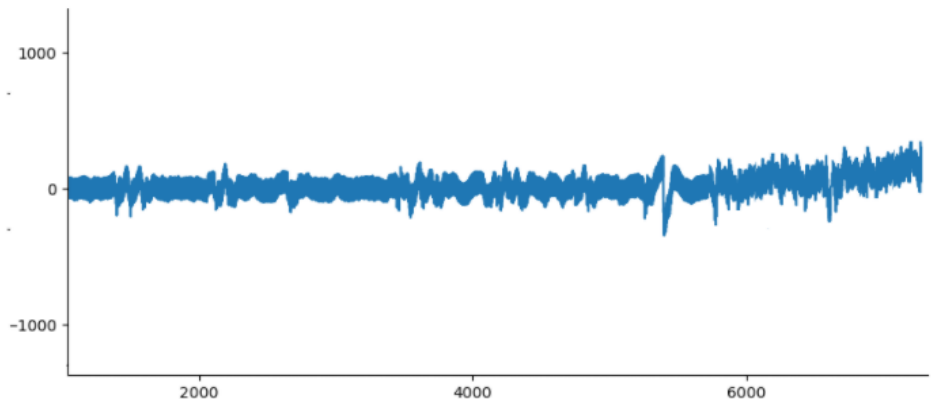

Fig.7. Alpha waves detection in EEG signal without low-pass filter

**Conclusion and discussion**
This is currently one of the easiest devices to use for reading and processing biosignals. This board is especially useful for researchers who set goals for controlling external objects, mechanical limbs through blinking, paradigms, and the image motor, and more. We will update the site, upload datasets and continue to improve the software and will publish cases on the use of this device to control robotic objects.
In the next version of the hardware, we will consider the possibility of installing gyroscopes and accelerometers to control the position of the object and add shields to protect against external electromagnetic interference.
During the development, we largely followed the interests of the user, specially choosing the most popular tools for signal processing. In the future, we plan to provide this function in the software for detecting the paradigms P300 and SSVEP. These paradigms are currently one of the most popular tools for controlling external objects using the power of thought. Our final task is to control a mechanical limb by motor imagery methods, which has not yet been fully realized through non-invasive methods. This method is the most natural and therefore the achievement of a positive result in this area will significantly improve the position of BCI in the world of neurobiology.

**Conflicts of Interest**: None
**Funding**: None
**Ethical Approval**: Not required